\documentclass[sigconf, nonacm]{acmart} %anonymous=true %[sigconf,review]

\begin{document}

%
% The "title" command has an optional parameter, allowing the author to define a "short title" to be used in page headers.
\title{Deploying a sharded MongoDB cluster as a queued job on a shared HPC architecture}

%
% The "author" command and its associated commands are used to define the authors and their affiliations.
% Of note is the shared affiliation of the first two authors, and the "authornote" and "authornotemark" commands
% used to denote shared contribution to the research.
\author{Aaron Saxton}
\email{saxton@illinois.edu}
\affiliation{%
  \institution{National Center For  Super Computing Applications \newline Blue Waters Project Office}
  \city{Urbana}
  \state{Illinois}
}

\author{Stephen Squaire}
\email{squaire3@illinois.edu}
\affiliation{%
  \institution{National Center For Super Computing Applications \newline Integrated Cyber Infrastructure}
  \city{Urbana}
  \state{Illinois}
}

%
% By default, the full list of authors will be used in the page headers. Often, this list is too long, and will overlap
% other information printed in the page headers. This command allows the author to define a more concise list
% of authors' names for this purpose.
\renewcommand{\shortauthors}{Saxton and Squaire}

%
% The abstract is a short summary of the work to be presented in the article.
\begin{abstract}
Data stores are the foundation on which data science, in all its variations, is built upon. They provide a queryable interface to structured and unstructured data. Data science often starts by leveraging these query features to perform initial data preparation. However, most data stores are designed to run continuously to service disparate user requests with little or no downtime. Many HPC architectures process user requests by job queue scheduler and maintain a shard filesystem to store a jobs persistent data. We deploy a MongoDB sharded cluster with a run script that is designed to run a data science workload concurrently. As our test piece, we run data ingest and data queries to measure the performance with different configurations on the Blue Waters supper computer.
\end{abstract}



%
% Keywords. The author(s) should pick words that accurately describe the work being
% presented. Separate the keywords with commas.

%
% This command processes the author and affiliation and title information and builds
% the first part of the formatted document.
\maketitle
\section{Introduction}
Queryable data stores have been a crucial part of modern data science. They provide a relatively small set of primitive operations by which a large variety of data analysis can be performed. Machine learning and other sophisticated statistical analysis are becoming more popular across all disciplines \cite{WilsonAS16}. These algorithms require more flexibility than the average SQL or NoSQL data store and can take advantage of a wider variety of compute resources than the average datastore host architecture. However, it is common for a data scientist to computationally balance their process between a datastore and compute resource. Data and algorithms are becoming large enough and computationally intensive enough that they can take advantage of a high performance architecture. The 2018 Gordon Bell prize "Employing Deep Learning Methods to Understand Weather Patterns" is an example. In this paper we present a performance profile of the NoSQL distributed data store MongoDB running as transient queued batch job on Blue Waters super computer. This will provide guidance about how to balance tasks in a data science pipelines on a HPC architectures.
\section{Background}
The major primitive operations of a traditional SQL data store are CRUD (Create, Read, Update, Delete), groupby, and join. NoSQL data stores, such as MongoDB, still maintain a core set of primitives that are analogous to traditional SQL: insert, find, update, remove, aggregate, lookup. This work will not examine all of these primitives, but we still wish to speak of them.

Finds and inserts naturally lend themselves to distributed computing. A program can dispatch them to many workers witch report back either nothing is found or the results of their task. The more sophisticated aggregate and lookup can not in general be executed without some synchronization or interprocess communication. Take for example a SQL inner join on column $C$ between tables $T_A$ and $T_B$. The naive algorithm goes as: Take first item in $C$ of table $T_A$, $C_{A0}$, perform a find on column $C$ in table $T_B$. In a distributed datastore, this will require each worker to look at $C_{AN}$ of $T_A$ hosted locally, search through the portion of $T_B$ hosted locally and pass $C_{AN}$ to every other worker. If the columns are indexed, there is some optimization you can do, but every worker will have to touch every other work at some point. Groupby haves similar problems. A map-reduce execution model can provide a single stage inter-worker communication to perform a groupby. The algorithms for performing joins and groupbys on a distributed data store are rebranded to give the user more flexibility. As such, this paper will only be concerned with insert and find type operations.

\section{System Setup}
Blue Waters is a mixture of Cray XE and XK blades with 4 Interlogos AMD processors XE nodes and two Interlogs AMD, K20 Nvidia GPU pairs on XK nodes. These nodes are all connected by Cray's Gemini interconnect. Shared filesystem storage is hosted on Cray's Synexion racks with a luster file system. Blue Waters has 4 login nodes that users can use to access the shared file system and submit run scripts for Moab and Torque schedule on the requested compute resources. \cite{Bode2013} \cite{Kramer2015}

Data stores are often intended to be persistent and service tasks that demand zero downtime. For instance, users data on the backend of a web app, hospital patient data, or product order data from a retailer. Compute resources on an HPC system that are available to a typical user are often ephemeral and are contrary to the execution model of common data stores. Fortunately, the final destination of a data stores data is on an underlying filesystem. MongoDB by default uses the WiredTiger storage engine to manage data underlying files on the host filesystem. For Blue Waters, that is a massively parallel and redundant file system\cite{Bode2013} \cite{Kramer2015}.
\subsection{MongoDB Basics}
A sharded MongoDB cluster runs with 3 types of worker processes that we host one on each processing element of a Blue Waters job: configuration server, shard server, and router. The configuration server manages the metadata of the collections \cite{MongoManShard}. Config servers store the metadata for a sharded cluster. The metadata reflects state and organization for all data and components within the sharded cluster. The metadata includes the list of chunks on every shard and the ranges that define the chunks\cite{MongoManConfigServ}. A shard contains a subset of sharded data for a sharded cluster. Together, the cluster?s shards hold the entire data set for the cluster\cite{MongoManShardServ}. Each shard will be assigned a unique filesystem path to deposit data too. MongoDB routers route queries and write operations to shards in a sharded cluster. A router provide the only interface to a sharded cluster from the perspective of applications. Applications never connect or communicate directly with the shards\cite{MongoManRouter}.
\begin{figure}[h]
  \centering
  \includegraphics[width=\linewidth]{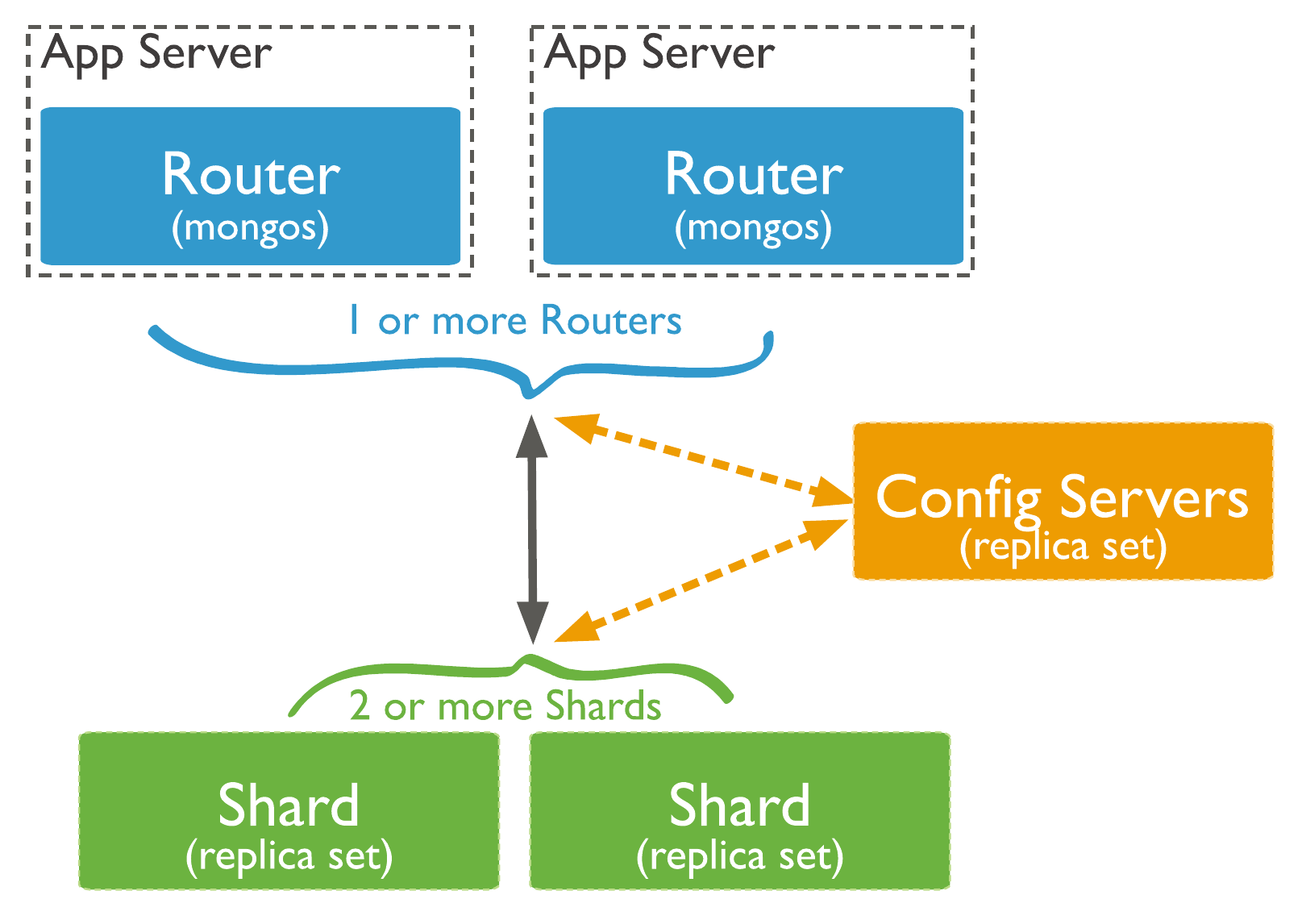}
  \caption{Sharded MongoDB cluster diagram \cite{MongoManShard}}
\end{figure}
\subsection{User Execution Model}
For a typical Blue Waters user to deploy a MongoDB cluster, they must construct a run-script that assigns to each processing element which roll it will be taking (config, shard, router). MongoDB is nativly deployed on a TCP/IP network therefore worker reference each other by hostname and port. In our run script we chose the number of processing elements to be equal number of node hosts. Once the assignments have been made, the run scrip will configure each worker to reference the config server. The config server will then share the settings with all members of the cluster. In addition to taking advantage of distributed compute resources, a sharded  MongoDB cluster will natively leverage the distributed nature of Luster on a Cray Synexian. When each shard worker is assigned a directory to place files, luster will distribute those files to an object storage server that should optimize further I/O.

This describes a run scripts execution for the datastore portion. In practice, once the datastore is active, a run script will continue making queries and processing. The runsript makes available through environment variables or a shared file a list of host names of the MongoBD clusters router servers. With this list, a run script may use either the mongo shell command or the Python package pymongo as the API to perform queries.

\section{Performance Data}
We perform 2 tasks, ingest data with insertMany with ordered=False and a conditional find on two indexed fields. These tasks are written in python using pymongo. The data we choose to ingest was time series metric data of Blue Waters compute nodes collected by OVIS. The data spans 5 years, sample each node independently once every minute, and includes about 75 distinct metrics (e.g. memory use, cpu activity, network activity). The totality of this data is about 70 billion rows. Storage for this data in flat csv file on Blue Waters Luster filesystem is about 200 terabytes. We index on timestamp and node id.

\begin{figure}[h]
  \centering
  \includegraphics[width=\linewidth]{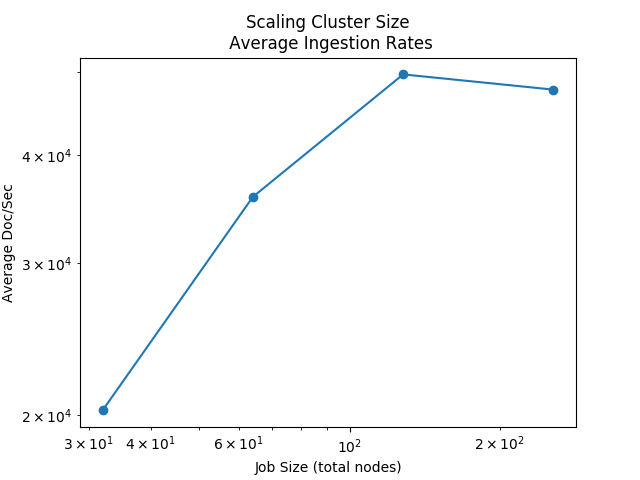}
  \caption{}
  \label{ingestScaling}
\end{figure}

We perform scaling analysis by increasing the number of shard-router pairs. For example, a job of 32 nodes is scheduled. 2 nodes will be for the configuration server, 7 shards, and 7 routers. This leaves 16 nodes to run the ingest script. Ingest is run with 4 processing elements per node, thus 64 insertMany will be processed concurrently  across 7 MongoDB routers. A job of 64 nodes would have 2 for configuration, 15 shards, 15 router servers and so on. The larger the cluster, the more data we upload. 
\begin{center}
\begin{table}[ht]
\caption{}
 \begin{tabular}{| c | c |} 

 \hline
 Nodes & Days of Data \\
 \hline
 32 & 3 \\
 64 & 7 \\
 128 & 14 \\
 256 & 14 \\
 \hline
\end{tabular}
 \label{daysTable}
\end{table}
\end{center}
A instertMany is performed by collecting a list of python dictionaries from the metric data csv file. The keys of the dictionaries become analogous to SQL table column names. A list of dictionaries is allowed to have different keys. However, ingesting this dataset allows the same keys for each document. Indexes were created for the timestamp key and node id key. In Figure  \ref{ingestScaling} MongoDB scales close to linear between 32, 64, and 128 nodes. We are still investigating the limitations at 256 nodes and beyond.

\begin{figure}[h]
  \centering
  \includegraphics[width=\linewidth]{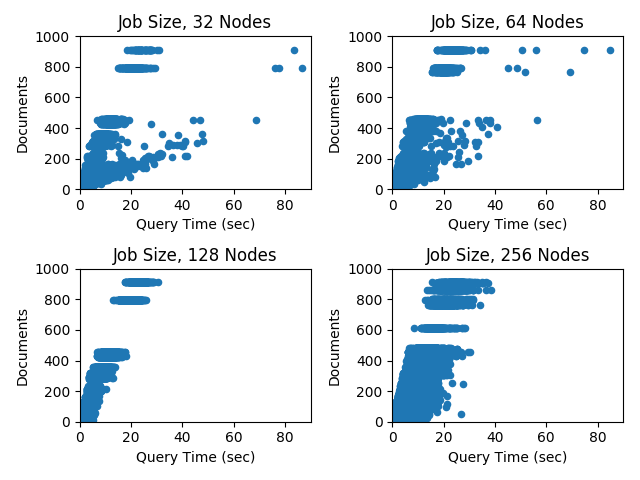}
  \caption{Sharded MongoDB cluster diagram \cite{MongoManShard}}
   \label{queryScaling}
\end{figure}

The query test was done by doing a conditional find. The query is constructed by reading user jobs metadata for time run, duration, and which nodes were assigned. The candidate user jobs were selected from a time period starting January 1, 2018 until the number of days described in Table \ref{daysTable}. The total number of documents returned by a query is number of user job nodes times duration of user job in minutes. Indeed, Figure \ref{queryScaling} indicates cluster size maintains a similar query performance for various MongoDB cluster sizes. It is important to point out that each cluster size is servicing more concurrent quarries. Cluster with size 32 was servicing between 16 and 64 concurrent find queries, cluster size 64 was between 32 and 128 queries, and so on.

%
% The acknowledgments section is defined using the "acks" environment (and NOT an unnumbered section). This ensures
% the proper identification of the section in the article metadata, and the consistent spelling of the heading.
\begin{acks}
This research is part of the Blue Waters sustained-petascale computing project, which is supported by the National Science Foundation (awards OCI-0725070 and ACI-1238993) and the state of Illinois. Blue Waters is a joint effort of the University of Illinois at Urbana-Champaign and its National Center for Supercomputing Applications.
\end{acks}

%
% The next two lines define the bibliography style to be used, and the bibliography file.
%\bibliographystyle{ACM-Reference-Format}
\bibliography{mongoHPC,bw}


\begin{thebibliography}{7}

%%% the Web address.

\ifx \showCODEN    \undefined \def \showCODEN     #1{\unskip}     \fi
\ifx \showDOI      \undefined \def \showDOI       #1{#1}\fi
\ifx \showISBNx    \undefined \def \showISBNx     #1{\unskip}     \fi
\ifx \showISBNxiii \undefined \def \showISBNxiii  #1{\unskip}     \fi
\ifx \showISSN     \undefined \def \showISSN      #1{\unskip}     \fi
\ifx \showLCCN     \undefined \def \showLCCN      #1{\unskip}     \fi
\ifx \shownote     \undefined \def \shownote      #1{#1}          \fi
\ifx \showarticletitle \undefined \def \showarticletitle #1{#1}   \fi
\ifx \showURL      \undefined \def \showURL       {\relax}        \fi
% invisible to TeX
\providecommand\bibfield[2]{#2}
\providecommand\bibinfo[2]{#2}
\providecommand\natexlab[1]{#1}
\providecommand\showeprint[2][]{arXiv:#2}

\bibitem[\protect\citeauthoryear{??}{Mon}{[n. d.]a}]%
        {MongoManConfigServ}
 \bibinfo{year}{[n. d.]}\natexlab{a}.
\newblock \bibinfo{title}{MongoDB Manual, Config Server}.
\newblock
\newblock
\urldef\tempurl%
\url{https://docs.mongodb.com/manual/core/sharded-cluster-config-servers/}
\showURL{%
Retrieved April 9, 2019 from \tempurl}


\bibitem[\protect\citeauthoryear{??}{Mon}{[n. d.]b}]%
        {MongoManRouter}
 \bibinfo{year}{[n. d.]}\natexlab{b}.
\newblock \bibinfo{title}{MongoDB Manual, Mongos}.
\newblock
\newblock
\urldef\tempurl%
\url{https://docs.mongodb.com/manual/core/sharded-cluster-query-router/}
\showURL{%
Retrieved April 9, 2019 from \tempurl}


\bibitem[\protect\citeauthoryear{??}{Mon}{[n. d.]c}]%
        {MongoManShardServ}
 \bibinfo{year}{[n. d.]}\natexlab{c}.
\newblock \bibinfo{title}{MongoDB Manual, Shard}.
\newblock
\newblock
\urldef\tempurl%
\url{https://docs.mongodb.com/manual/core/sharded-cluster-shards/}
\showURL{%
Retrieved April 9, 2019 from \tempurl}


\bibitem[\protect\citeauthoryear{??}{Mon}{[n. d.]d}]%
        {MongoManShard}
 \bibinfo{year}{[n. d.]}\natexlab{d}.
\newblock \bibinfo{title}{MongoDB Manual, Sharding}.
\newblock
\newblock
\urldef\tempurl%
\url{https://docs.mongodb.com/manual/sharding/}
\showURL{%
Retrieved April 9, 2019 from \tempurl}


\bibitem[\protect\citeauthoryear{Bode, Butler, Dunning, Hoefler, Kramer, Gropp,
  and Hwu}{Bode et~al\mbox{.}}{2013}]%
        {Bode2013}
\bibfield{author}{\bibinfo{person}{Brett Bode}, \bibinfo{person}{Michelle
  Butler}, \bibinfo{person}{Thom Dunning}, \bibinfo{person}{Torsten Hoefler},
  \bibinfo{person}{William Kramer}, \bibinfo{person}{William Gropp}, {and}
  \bibinfo{person}{Wen-mei Hwu}.} \bibinfo{year}{2013}\natexlab{}.
\newblock \showarticletitle{The {B}lue {W}aters Super-System for
  Super-Science}.
\newblock In \bibinfo{booktitle}{\emph{Contemporary High Performance
  Computing}}. \bibinfo{publisher}{Chapman and Hall/CRC},
  \bibinfo{pages}{339--366}.
\newblock
\showISBNx{978-1-4665-6834-1}
\urldef\tempurl%
\url{https://www.taylorfrancis.com/books/e/9781466568358}
\showURL{%
\tempurl}


\bibitem[\protect\citeauthoryear{Kramer, Butler, Bauer, Chadalavada, and
  Mendes}{Kramer et~al\mbox{.}}{2015}]%
        {Kramer2015}
\bibfield{author}{\bibinfo{person}{William Kramer}, \bibinfo{person}{Michelle
  Butler}, \bibinfo{person}{Gregory Bauer}, \bibinfo{person}{Kalyana
  Chadalavada}, {and} \bibinfo{person}{Celso Mendes}.}
  \bibinfo{year}{2015}\natexlab{}.
\newblock \showarticletitle{{Blue Waters Parallel I/O Storage Sub-system}}.
\newblock In \bibinfo{booktitle}{\emph{High Performance Parallel I/O}},
  \bibfield{editor}{\bibinfo{person}{Prabhat} {and} \bibinfo{person}{Quincey
  Koziol}} (Eds.). \bibinfo{publisher}{CRC Publications, Taylor and Francis
  Group}, \bibinfo{pages}{17--32}.
\newblock
\showISBNx{978-1-4665-8234-7}


\bibitem[\protect\citeauthoryear{Wilson, Alter, and Shukla}{Wilson
  et~al\mbox{.}}{2016}]%
        {WilsonAS16}
\bibfield{author}{\bibinfo{person}{H.~James Wilson}, \bibinfo{person}{Allan
  Alter}, {and} \bibinfo{person}{Prashant Shukla}.}
  \bibinfo{year}{2016}\natexlab{}.
\newblock \bibinfo{title}{Companies Are Reimagining Business Processes with
  Algorithms}.
\newblock
\newblock
\urldef\tempurl%
\url{https://hbr.org/2016/02/companies-are-reimagining-business-processes-with-algorithms}
\showURL{%
Retrieved April 9, 2019 from \tempurl}


\end{thebibliography}
% 
% If your work has an appendix, this is the place to put it.

\end{document}